\documentclass[showkeys,nofootinbib]{revtex4}

\usepackage{amsfonts}
\usepackage{amssymb}
\usepackage{natbib}
\usepackage[latin1]{inputenc}
\usepackage[babel]{csquotes}
\usepackage{graphicx}
\usepackage[english]{babel}

\usepackage[T1]{fontenc}
\usepackage[centertags]{amsmath}

\usepackage{mathrsfs}
\usepackage{amsmath}
\usepackage{amsthm}
\usepackage{amssymb}
\usepackage{mathtools}
\usepackage{wasysym}
\usepackage{subfigure}
\usepackage{multirow}

\DeclareFontFamily{U}{mathb}{\hyphenchar\font45}
\DeclareFontShape{U}{mathb}{m}{n}{
      <5> <6> <7> <8> <9> <10> gen * mathb
      <10.95> mathb10 <12> <14.4> <17.28> <20.74> <24.88> mathb12
      }{}
\DeclareSymbolFont{mathb}{U}{mathb}{m}{n}

\DeclareMathSymbol{\Sun}{3}{mathb}{"40}

\begin{document}
\title{Perturbative and semi-analytical solutions to Teukolsky equations for massive fermions}
\author{Mattia Villani}
\affiliation{DISPEA, Universit\`a di Urbino Carlo Bo via Santa Chiara, 27 61029, Urbino, Italy}
\affiliation{INFN - Sezione di Firenze via B.Rossi, 1 50019, Sesto Fiorentino, Florence, Italy}
\email{mattia.villani@uniurb.it}
\date{\today}

\begin{abstract}
In this work, we aim at solving the Teukolsky equations for a fermion with mass $m_e\neq 0$ in the presence of a rotating black hole with mass $M$. We consider two different regimes: $\tilde{m}_e= M^{-1} m_e\ll 1$ and $a\omega \ll 1$; $\tilde{m}_e\ll 1$ and $a\omega \gtrsim 1$. We treat each of these two regimes in different ways: we use a perturbative approach for the first, similar to the \emph{usual} one employed for spin 0, 1, 2 and mass-less 1/2 fields, but with two small parameters (a$\omega$ and $\tilde{m}_e$); as we shall see, the second can be treated with a semi-analytical approach. In a forthcoming paper we shall study the remaining two cases in which $\tilde{m}_e \gtrsim 1$, while $a\omega \ll 1$ or $a\omega \gtrsim 1$. The regime with $\tilde{m}_e \ll 1$, but $a\omega\gtrsim 1$ is probably the most interesting from the astrophysics point of view, but this last two cases might be of some interest for the study of the interaction of fermions with very small black holes, which may be formed, for example, in the last stages of the Hawking evaporation.
\end{abstract}
\maketitle

\section{Introduction}
Gravitational waves (GW) in General Relativity were first predicted by Einstein soon after the publication of the gravitational field equations in 1915 (see the review \cite{histo1,histo2}). They are ripples in the geometry of spacetime propagating at the speed of light. GW can be generated in many ways, most notably by the merging of binary systems constituted by two black holes or two neutron stars or of \emph{mixed} binaries constituted by a black hole and a neutron star. The first indirect proof of the existence of gravitational waves has been obtained with the study of the change in the revolution period of the Hulse-Taylor binary pulsar, first discovered in 1975 \cite{HTp}: the measured change of the revolution period of the binary due to the loss of energy and angular momentum from the emission of GWs and the predictions of General Relativity are in astonishing agreement \cite{HTp2}; for this work, Hulse and Taylor were awarded the Noble prize in Physics in 1993. The efforts for the direct detection of GW started back in the '70s with resonant mass experiments and culminated with the detection of the merging of two black holes, the event named GW140915, by the ground-based interferometers LIGO and Virgo \cite{GWdet1,GWdet2,GWdet3}. Since then, many merging events were observed, most notably the event GW170817, the merging of two neutron stars, which also produced a short Gamma-Ray Burst, GRB170817A. This event marks the beginning of the multi-messenger astronomy era \cite{2NS}. In future, GWs will be a major tool for the study of fundamental Physics, with applications ranging from astrophysics and cosmology to nuclear and sub-nuclear physics \cite{histo2,fut1,fut2,fut3,fut4,fut5}.

Black holes can generate GWs, but also scatter and absorb them. The calculation of the scattering and absorbtion cross-sections started with the works by Teukolsky \cite{Teu1} and Press \cite{Teu2}. Many works have been published since then, dealing with scalar fields \cite{Scal}, spin 1 fields \cite{Vect,Vect2,Vect3,Vect4,Vect5} and also spin 2 fields \cite{SD}. The theoretical and mathematical work has been fundamental \cite{Matz,Matz2,Matz3,Matz4,SN,MST}, see also the review \cite{lrr}. 

Spin 1/2 fermions can also be scattered and absorbed by black holes. Works on this topic have focused mainly on mass-less fermions, i.e. neutrinos, \cite{Ferm}. Fermions are, however, massive particles (and also some, if not all, of neutrino flavors must have mass, albeit very small, see \cite{numass1,numass2,numass3}). The presence of a mass term, $m_e\neq 0$, unfortunately, makes the Teukolsky equations for fermions much more complicated and the \emph{usual} techniques (see for example \cite{lrr,libro}) cannot be applied, since the equations do not constitute an eigenvalue problem anymore. We present here a possible way to tackle this problem.

In this paper, we identify four different regimes which can be treated in different ways: in the first regime, we have both $\tilde{m}_e = m_e \, M^{-1}\ll 1$ (where $M$ is the black hole mass) and $a\omega\ll 1$: this can be treated perturbatively in a way similar to the \emph{usual} one, but with two expansion parameters: $a\omega$ and $\tilde{m}_e$; in the second regime we have, on the contrary, $\tilde{m}_e\gtrsim 1$ and $a\omega\gtrsim 1$: no analytical treatment is possible in this case, so this must be solved numerically; in the third regime, we have $\tilde{m}_e\ll 1$, but $a\omega\gtrsim 1$: this can be treated in a semi-analytical way, by adding analytical perturbative corrections to the mass-less solution. There is also a fourth case in which $\tilde{m}_e\gtrsim 1$, while $a\omega\ll 1$, which will be treated in a separated, forthcoming paper.

This paper is organized as follows: in Section \ref{sec:teu} we present Teukolsky equations for massive fermions; in Section \ref{sec:small} we present the treatment of the angular Teukolsky equation for the first regime with $m_e\ll 1$ and $a\omega\ll 1$; in Section \ref{sec:semi-an} the semi-analytical treatment for $m_e\ll 1$ and $a\omega\gtrsim 1$. The treatment of the radial Teukolsky equation is presented in detail in Sections \ref{sec:radial} in which we find the general form of the solution and we shall also find that the resulting equation is very reminiscent of a Heun differential equation: this observation will guide us through the (somehow cumbersome) mathematical analysis. We shall expand the solution in terms of Gauss' Hypergeometric functions ${}_2F_1(a,b,c;x)$ and calculate the renormalized angular momentum up to second perturbative order for the regimes $\tilde{m}_e\ll 1$, $a\omega\ll 1$ and $\tilde{m}_e\ll 1$, $a\omega\gtrsim 1$. The recurrence relation found in this case is not of the three-terms type (we shall see that it has in fact seven terms), so we also review the mathematical theory behind these higer order recurrence relations and their link to n-continued fractions (a generalization of continued fractions with several terms). In Section \ref{sec:conclusion} we present our conclusions. There is also an appendix, Appendix \ref{app:uno}, in which we list the expressions for the coefficients of the recurrence relation.

\section{Teukolsky equations for massive fermion}
\label{sec:teu}

The radial Teukolsky equation for a massive fermion in the Kerr spacetime is given by \cite{libro}:
\begin{equation}\label{eq:radial}
\begin{split}
&\Delta^{-s}\frac{d}{dr} \left( \Delta^{s+1} \frac{d\;{}_sR(r)}{dr}\right) + \left( \frac{K^2-2i s(r-M)K}{\Delta} +4 i s \omega r - \lambda \right) \; {}_sR(r)+\\
&-\left[ \left( \frac{2i s M \tilde{m}_e \Delta}{\sqrt{\lambda}-2i s M \tilde{m}_e r} \right) \left( \frac{d}{dr} -\frac{2i s K}{\Delta}+\frac{(2s+1)(r-M)}{2\Delta} \right) -M \tilde{m}_e^2r^2 \right]\;{}_sR(r)=0
\end{split}
\end{equation}
while the angular equation is given by \cite{libro}:
\begin{equation}\label{eq:angular}
\begin{split}
&\frac{1}{\sin(\theta)} \frac{d}{d\theta}\left( \sin(\theta)\frac{d\; {}_sS_l^m(\theta,\phi)}{d\theta} \right)+\\
+&\left( a^2\omega^2\cos^2(\theta)-\frac{m^2}{\sin^2(\theta)}-2a\omega s \cos(\theta)-\frac{2m s \cos(\theta)}{\sin^2(\theta)}-s^2\cot^2(\theta)+\lambda -a^2\omega^2+2a m \omega+s \right) {}_sS_m^l(\theta,\phi)+\\
&\left( \frac{a M \tilde{m}_e\sin(\theta)}{\sqrt{\lambda}-2aM\tilde{m}_e s \cos(\theta)} \right) \left( \frac{d}{d\theta} - 2 a m \omega s \sin(\theta)+\frac{2m s}{\sin(\theta)} - a^2 M^2 \tilde{m}_e^2 \cos^2(\theta) \right) {}_sS_m^l(\theta,\phi)=0,
\end{split}
\end{equation}
if the spin $s$ is negative ($s=-1/2$); if $s=1/2$, in the second line of \eqref{eq:radial} and in the third line of \eqref{eq:angular} we have to make the substitution:
\begin{equation}
\sqrt{\lambda} \mapsto \sqrt{\lambda+1}.
\end{equation}
In the above formulas, we have \cite{lrr,libro}:
\begin{equation}
\Delta=r^2-2 M r +a^2 \quad K=(r^2+a^2)\omega-a m \quad \tilde{m}_e=\frac{m_e}{M},
\end{equation}
while $m_e$ is the fermion mass and $\lambda$ is a separation constant, $a$ and $M$ are the black hole angular momentum and mass and $\omega$ is the frequency of the incoming radiation.

The presence of the mass term complicates the solution of the equations, since in this case $\lambda$ cannot be found simply by solving an eigenvalue problem as usually done for spin 0, 1 or 2 fields or for mass-less fermions. We can, however, recognize four different regimes according to the values of $a\omega$ and $\tilde{m}_e$:
\begin{enumerate}
\item Both $a\omega$ and $\tilde{m}_e$ are \emph{small}: this case can be treated pertubatively in a similar way to what is usually done for other values of the spin, but, in this case, there will be two expansion parameters;\label{itm:small}
\item Both $a\omega$ and $\tilde{m}_e$ are \emph{large}: this requires a full-fledged numerical approach. We shall treat this regime in a forthcoming paper;\label{itm:large}
\item In the third regime, we have that $a\omega\gtrsim 1$, but still $\tilde{m}_e\ll 1$. This regime is the one that is probably the most interesting from the astrophysics point of view, since for any black hole known $M \gg m_e$: in this case, the smallness of the ratio $\tilde{m}_e$ allows for a perturbative semi-analytical approach, i.e. we can obtain analytical corrections due to the presence of $m_e$ and add them to the numerical solution of the Teukolsky equations for a finite value of $a\omega$;\label{itm:smallarge}
\item There is also a fourth regime, in which $a\omega \ll 1$, while $\tilde{m}_e\gtrsim 1$. In this case, the black hole mass $M$ is smaller than the fermion mass $m_e$. This regime might be important for the study of microscopic black holes originated, for example, from the last stages of the Hawking evaporation process, and their interaction with fundamental particles. This regime can also be treated in a semi-analytical way, similarly to the third regime, starting from the numerical solution of the equations for $a\omega=0$ but $\tilde{m}_e\neq 0$, and then applying analytical corrections due to the finiteness of $a\omega$. This case will be treated separately in a forthcoming, dedicated paper.
\end{enumerate}

In the following section, we first tackle the problem of solving the angular equation in the regime \ref{itm:small} and then we present our semi-analytical approach for regime \ref{itm:smallarge}.

\section{Regime $a\omega\ll 1$, $\tilde{m}_e\ll 1$ - Approximate analytical solution of the angular equation}
\label{sec:small}
In order to solve the angular equation \eqref{eq:angular}: we use a perturbative approach, that is, we assume that both the mass of the fermion $\tilde{m}_e$ and the parameter $a\omega$ are \emph{small}, so we can use a double expansion in $\tilde{m}_e$ and $a\omega$. In the following we discuss in some details the calculations for the case of negative spin $s$: the case for a positive spin can be treated similarly, so we simply show the results.

We start by expanding the solution ${}_sS^m_l(\theta)$ in series up to second order in $a\omega$ and $\tilde{m}_e$:
\begin{equation}\label{eq:exp-S}
\begin{split}
{}_sS^m_l(\theta,\phi) &= S_{(0,0)}^{lms}(\theta,\phi) + a\omega S_{(1,0)}^{lms}(\theta,\phi) + \tilde{m}_e S_{(0,1)}^{lms}(\theta,\phi) +\\
&+ a^2\omega^2 S_{(2,0)}^{lms}(\theta,\phi) + \tilde{m}_e^2 S_{(0,2)}^{lms}(\theta,\phi) + \tilde{m}_e a\omega S_{(1,1)}^{lms}(\theta,\phi) + O(m_e^2,a^2\omega^2).
\end{split}
\end{equation}
Similarly, for the separation constant $\lambda$ we write:
\begin{equation}\label{eq:exp-l}
\lambda= \lambda_{(0,0)} + a\omega \lambda_{(1,0)} + \tilde{m}_e \lambda_{(0,1)} + a^2\omega^2 \lambda_{(2,0)} + \tilde{m}_e^2 \lambda_{(0,2)} + a\omega \tilde{m}_e \lambda_{(1,1)}  + O(m_e^2,a\omega^2).
\end{equation}
We use the notation:
\begin{equation*}
S_{(i,j)}^{lms}(\theta,\phi)
\end{equation*}
by which we mean that the expression is at order $O(a^i\omega^i,m_e^j)$. The notation for the separation constant is similar.

We now substitute \eqref{eq:exp-S} and \eqref{eq:exp-l} in \eqref{eq:angular} and separate the various contributions by grouping terms of the same order, thus we find:
\begin{enumerate}
\item At zeroth order, we get the equation:
\begin{equation}\label{eq:zero}
\frac{1}{\sin(\theta)} \frac{d}{d\theta} \left( \sin(\theta) \frac{d\;S_{(0,0)}^{lms}(\theta,\phi)}{d\theta} \right) + \left[ \left(s \cot(\theta) + \frac{1}{\sin(\theta)}\right) + \lambda_{(0,0)} +s \right] S_{(0,0)}^{lms}(\theta,\phi) = 0.
\end{equation}
\item At the first order we have:
\begin{equation}\label{eq:first}
2a\omega \left( m-s\cos(\theta) \right) S_{(0,0)}^{lms}(\theta,\phi) + \frac{M\,\tilde{m}_e}{\lambda_{(0,0)}} \left( 2a m s+a \sin(\theta) \frac{d\;S_{(0,0)}^{lms}(\theta,\phi)}{d\theta}\right).
\end{equation}
\item At second order, we have:
\begin{equation}\label{eq:second}
\begin{split}
&a\omega^2 \left[ 2(m -s \cos(\theta)) S_{(1,0)}^{lms}(\theta,\phi) - \sin^2(\theta) S^{lms}_{(0,0)}(\theta,\phi)  \right] +\\
+ &\tilde{m}_e^2 \Bigg[ M^2\, \left( \frac{4a^2m s^2}{\lambda_{(0,0)}} \cos(\theta) + \frac{2a^2}{\lambda_{(0,0)}} \frac{d}{d\theta} \right) \, S_{(0,0)}^{lms}(\theta,\phi) + \frac{2Mams}{\sqrt{ \lambda_{(0,0)}}} S_{(0,1)}^{lms}(\theta,\phi) +\\
&- \frac{M \, \lambda_{(0,1)}}{\lambda_{(0,0)}^{3/2}} \left( a m s+ a\sin(\theta) \frac{d}{d\theta} \right) S_{(0,0)}^{lms}(\theta,\phi) \Bigg] +\\
+ &M\,\tilde{m}_e a\omega \Bigg[ 2(m-s\cot(\theta)) S_{(0,1)}^{lms} + \frac{2ams}{\sqrt{\lambda_{(0,0)}}}  S_{(1,0)}^{lms}(\theta,\phi) +\\
&- \left( \frac{2ams}{\sqrt{\lambda_{(0,0)}}} \sin^2(\theta) + \frac{\lambda_{(1,0)}}{\lambda^{3/2}_{(0,0)}} \left( am+\frac{1}{2}a\sin(\theta)\frac{d}{d\theta} \right) \right) S_{(0,0)}^{lms}(\theta,\phi) \Bigg].
\end{split}
\end{equation}
\end{enumerate}
In the following subsections we shall calculate the perturbations to the eigenvalue $\lambda$ at the various perturbative orders with equations \eqref{eq:zero}-\eqref{eq:second}.

\subsection{Zeroth order}
At this order, it is known \cite{lrr,SWSA} that:
\begin{equation}
\lambda_{(0,0)}=l(l+1)-s(s+1)
\end{equation}
and that the functions $S_{(0,0)}^{lms}$ are spin-weighted spherical harmonics ${}_sY_{lm}(\theta,\phi)$, \cite{SWSA}.

\subsection{First order}
At this order there are two corrections, $\lambda_{(1,0)}$ and $\lambda_{(0,1)}$, which are given by:
\begin{align}
\lambda_{(1,0)} &=\int d\Omega \left({}_sY^{*}_{lm}(\theta,\phi)(2m-2s\cos(\theta)){}_{s_{1}}Y_{l_{1}m_{1}}(\theta,\phi)\right)\\
\lambda_{(0,1)} &=\int \frac{d\Omega}{\sqrt{\lambda_{(0,0)}}} \left({}_sY^{*}_{lm}(\theta,\phi)\left( 2M a m s+a M \sin(\theta) \frac{d}{d\theta}\right){}_{s_{1}}Y_{l_{1}m_{1}}(\theta,\phi)\right).
\end{align}
These integrals (and also those in the following subsections) can be calculated by writing the trigonometric functions as linear combination of spherical harmonics, or spin-weighted spherical harmonics and using the following properties of the spin-weighted spherical harmonics \cite{SWSA}:
\begin{equation}
{}_sY_{lm}(\theta,\phi) = \sqrt{\frac{2l+1}{4\pi} \frac{(l+m)!(l-m)!}{(l+s)!(l-s)!}} \sin^{2l}\left( \frac{\theta}{2} \right) \exp(i m \phi) \sum_{r=0}^{l-s} \binom{l-s}{r} \binom{l+s}{r+s-m} (-1)^{l-s-r} \cot^{2r+s-m}\left( \frac{\theta}{2} \right),
\end{equation}
where the first and second terms inside the sum in the right hand side of the above expression are binomial coefficients; we have also that:
\begin{align}\label{eq:pro1}
&{}_sY^{*}_{lm}(\theta,\phi) =(-1)^{s+m} \; {}_{-s}Y_{l-m}(\theta,\phi),\\
&\int d\Omega({}_sY^{*}_{lm}(\theta,\phi)\; {}_{s}Y_{l_{1}m_{1}}(\theta,\phi)) = \delta_{ll_{1}} \delta_{mm_{1}},\\\label{eq:pro2}
&\int d\Omega \left( {}_{s_{1}}Y_{l_{1}m_{1}}(\theta,\phi)\;{}_{s_{2}}Y_{l_{2}m_{2}}(\theta,\phi)\;{}_{s_{3}}Y_{l_{3}m_{3}}(\theta,\phi) \right) =\\\nonumber
&= \sqrt{\frac{(2l_1+1)(2l_2+1)(2l_3+1)}{4\pi}} \left( \begin{array}{ccc}
l_1 & l_2 & l_3\\
m_1 & m_2 & m_3
\end{array} \right)\left( \begin{array}{ccc}
l_1 & l_2 & l_3\\
-s_1 & -s_2 & -s_3
\end{array} \right),\\\label{eq:profin}
&{}_{s+1}Y_{lm}(\theta,\phi) = -\sin^s(\theta) \left[ \frac{d}{d\theta}\left(\frac{{}_sY_{lm}(\theta,\phi)}{\sin^s(\theta)}\right) - \frac{i}{\sin(\theta)} \frac{d}{d\phi}\left(\frac{{}_sY_{lm}(\theta,\phi)}{\sin^s(\theta)}\right) \right],
\end{align}
where
\begin{equation*}
\left( \begin{array}{ccc}
l_1 & l_2 & l_3\\
m_1 & m_2 & m_3
\end{array} \right)
\end{equation*}
are the three-j symbols and where the spins in the second three-j symbol in \eqref{eq:pro2} satisfy the relation $s_1+s_2+s_3=0$. As said above, we need to express the sine and cosine functions as linear combinations of (spin-weighted) spherical harmonics. We find the following relations:
\begin{equation*}
\cos(\theta) = \sqrt{\frac{4\pi}{3}} Y_{10}(\theta,\phi), \qquad \sin(\theta)= \sqrt{\frac{8\pi}{3}}\, {}_1Y_{1,0}(\theta,\phi).
\end{equation*}

Putting everything together, we find that:
\begin{align*}
\lambda_{(1,0)} &= -2m \left( 1+\frac{s^2}{l(l+1)} \right)\\
\lambda_{(0,1)} &= \frac{aMm}{\sqrt{\lambda_{(0,0)}}} \left( (2s-1) - \frac{s^2}{l(l+1)} \right).
\end{align*}

For the case of positive spin, one can show that in the last expression, the term in the square root must be modified as follows:
\begin{equation*}
\lambda_{(0,0)} \rightarrow \lambda_{(0,0)}+1
\end{equation*}

\subsection{Second order perturbation}
At this order there are three corrections, namely $\lambda_{(2,0)},\lambda_{(0,2)}$ and $\lambda_{(1,1)}$ which are calculated, respectively, by multiplying on the left each row in equation \eqref{eq:second} by ${}_sY_{lm}^{*}$ and integrating using the properties \eqref{eq:pro1}-\eqref{eq:profin}. There also appear integrals of the form:
\begin{equation*}
\int d\Omega({}_sY_{lm}^{*}(\theta,\phi)\, \mathcal{O}\, S_{(0,1)}^{lms}(\theta,\phi)), \qquad \int d\Omega({}_sY_{lm}^{*}(\theta,\phi)\, \mathcal{O}\, S_{(1,0)}^{lms}(\theta,\phi)).
\end{equation*}
where $\mathcal{O}$ is a generic operator involving trigonometric functions and derivatives. These integrals can be calculated by assuming that higher order functions $S^{lms}_{(0,1)},S^{lms}_{(1,0)}$ are given as linear combination of ${}_sY_{lm}$ as follows:
\begin{equation}
S_{(0,1)}^{l_{i}ms}(\theta,\phi) =  \sum_j c_{ij}\; {}_sY_{l_{j}m}(\theta,\phi) \qquad S_{(1,0)}^{l_{i}ms}(\theta,\phi) =  \sum_j c^\prime_{ij}\; {}_sY_{l_{j}m}(\theta,\phi)
\end{equation}
where:
\begin{equation}
c_{lm} =\frac{aM}{\sqrt{\lambda_{(0,0)}}} \; \frac{\int d\Omega\left[ {}_sY^{*}_{l_{i}m}(\theta,\phi) \left( 2ms + \sin(\theta) \frac{d}{d\theta} \right)^2 {}_sY_{l_{j}m}(\theta,\phi) \right]}{l(l+1)-k(k+1)}
\end{equation}
\begin{equation}
c^\prime_{lm} =\frac{\int d\Omega\left[ {}_sY^{*}_{l_{i}m}(\theta,\phi) \left( 2m-2s\cos(\theta) \right)^2 {}_sY_{l_{j}m}(\theta,\phi) \right]}{l(l+1)-k(k+1)}.
\end{equation}
With all this, we find the following expressions:
\begin{align}
\lambda_{(2,0)} &= H(l+1)-H(l)  \quad \text{with} \quad H(l)= \frac{(l^2-m^2)(l^2-s^2)^2}{(2l-1)l^3(2l+1)},\\\label{eq:questa}
\lambda_{(0,2)} &=-\frac{4a^2M^2ms^2}{\lambda_{(0,0)}} \frac{ms}{l+l^2}+\frac{a^2M^2s}{\lambda_{(0,0)}} \frac{s(-3m^2+(l^2+l)(-1+2l(l+1)+2m^2))-2s^2(l^2+2-3m^2)}{l(l+1)(2l-1)(3+2l)}+\\\nonumber
&-\frac{aMms \,\lambda_{(0,1)}}{\lambda_{(0,0)}^{3/2}}+\frac{aM\,\lambda_{(0,1)}}{2m\lambda_{(0,0)}^{3/2}} \left( \frac{s^2}{l^2+l}+1 \right) + c^{\prime}_{lm},\\\label{eq:quella}
\lambda_{(1,1)} &= - \frac{2Mams}{\lambda_{(0,0)}^{1/2}} \frac{s(-3m^2+(l^2+l)(-1+2l(l+1)+2m^2))-2s^2(l^2+2-3m^2)}{l(l+1)(2l-1)(3+2l)}\\\nonumber
&-\frac{aMms\, \lambda_{(0,1)}}{2\lambda_{(0,0)}^{3/2}}+\frac{aM\,\lambda_{(0,1)}}{2m\lambda_{(0,0)}^{3/2}} \left( \frac{s^2}{l+l^2} + 1 \right) + c_{lm}.
\end{align}

In order to obtain the above expressions, we have used the following results:
\begin{align*}
\cos^2(\theta) &= \sqrt{\frac{16\pi}{45}} Y_{20} + \sqrt{\frac{4\pi}{9}} Y_{00}\\
\sin^2(\theta) &= \sqrt{\frac{16\pi}{9}} Y_{00} - \sqrt{\frac{16\pi}{45}} Y_{20}\\
\cos(\theta)\sin(\theta) &= \sqrt{\frac{8\pi}{15}} {}_1Y_{20}
\end{align*}

As above, for the positive spin case one should carry out the following substitution in \eqref{eq:questa} and \eqref{eq:quella}:
\begin{equation*}
\lambda_{(0,0)} \rightarrow \lambda_{(0,0)} +1.
\end{equation*}

\section{Regime $a\omega\gtrsim 1$ $\tilde{m}_e\ll 1$ - A semi-analytical approach}
\label{sec:semi-an}

For the last regime we consider in this paper, number \ref{itm:smallarge} of section \ref{sec:teu}, with \emph{small} $\tilde{m}_e$ and \emph{large} $a\omega$, we can  assume that we can expand ${}_sS_{lm}(\theta,\phi)$ and $\lambda$ as follows:
\begin{align}\label{eq:expang2}
&{}_sS_{lm}(\theta,\phi)= {}_sS^{(0)}_{lm}(\theta,\phi) + \tilde{m}_e \, {}_sS^{(1)}_{lm}(\theta,\phi)+ \tilde{m}_e^2 \, {}_sS^{(2)}_{lm}(\theta,\phi)+O(m_e^3),\\\label{eq:explambda2}
&\lambda=\lambda_0 + \tilde{m}_e \, \lambda_1 + \tilde{m}_e^2 \, \lambda_2 + O(m_e^3)
\end{align}
where ${}_sS^{(0)}_{lm}(\theta,\phi)$ are spheroidal harmonics and ${}_sS^{(i)}_{lm}(\theta,\phi)$ for $i \geq 1$ are the pertubative corrections; similarly $\lambda_0$ is the zeroth order separation constant, which can be obtained from spectral method (see \cite{huges1,huges2}) and $\lambda_i$ are higher order perturbative corrections; finally, each of the ${}_sS_{lm}^{(i)}$ are expanded as a sum of spin-weighted spherical harmonics. We can now expand the angular equation up to the desired order in $\tilde{m}_e$. In this way, we find analytical corrections to the separation constant and to the coefficients of the expansion in spin-weighted spherical harmonics. This greatly simplifies the solution of the equation. 

By substituting \eqref{eq:expang2} and \eqref{eq:explambda2} into \eqref{eq:angular} and separating the various powers of $\tilde{m}_e$, we find:
\begin{enumerate}
\item At zeroth order we find the \emph{usual} spheroidal harmonics equation:
\begin{equation}\label{eq:angular2}
\begin{split}
&\frac{1}{\sin(\theta)} \frac{d}{d\theta}\left( \sin(\theta)\frac{d\; {}_sS^{(0)}_{lm}(\theta,\phi)}{d\theta} \right)+\\
+&\left( a^2\omega^2\cos^2(\theta)-\frac{m^2}{\sin^2(\theta)}-2a\omega s \cos(\theta)-\frac{2m s \cos(\theta)}{\sin^2(\theta)}-s^2\cot^2(\theta)+\lambda -a^2\omega^2+2a m \omega+s \right) {}_sS^{(0)}_{lm}(\theta,\phi)=0,
\end{split}
\end{equation}
\item At the first order we have:
\begin{equation}\label{eq:firstorder}
\frac{a \, M}{\sqrt{\lambda_{0}}} \sin(\theta) \frac{dS^{(0)}_{lm}(\theta,\phi)}{d\theta}+ \frac{2Mams}{\sqrt{\lambda_{0}}} \left( 1 -a\omega \, \sin^2(\theta) \right) \, {}_sS^{(0)}_{lm}(\theta,\phi)=0,
\end{equation}
\item At second order we have:
\begin{equation}\label{eq:secondorder}
\begin{split}
&\frac{aMms}{\lambda_{0}^{3/2}}\,\left[4aMs\, \lambda^{1/2}_{0} \,\cos(\theta) -  \lambda_{1} \right]\, \left( 1 - a\omega \, \sin^2(\theta) \right)\, {}_sS^{(0)}_{lm}(\theta,\phi) +\\
+&\frac{a}{\lambda_0^{3/2}} \, \sin(\theta)\,\left( 2\,aM^2s\,\lambda^{1/2}_{0} \cos(\theta) - \lambda_{1} \right) \, \frac{d{}_sS^{(0)}_{lm}(\theta,\phi)}{d\theta} + \frac{2Mams}{\lambda^{1/2}_{0}} \left( 1 -a\omega \, \sin^2(\theta) \right) \, {}_sS^{(1)}_{lm}(\theta,\phi) =0.
\end{split}
\end{equation}
\end{enumerate}

We now discuss the analytical corrections to the separation constant $\lambda$ in the following subsections.

\subsection{Zeroth order}
As said above, at the lowest order in the expansion in $\tilde{m}_e$, we obtain the \emph{usual} angular equation which can be treated numerically with the spectral method, thus obtaining the separation constant at zero order $\lambda_0$, see \cite{huges1,huges2}.

\subsection{First order}
At the first order, we substitute the expansion:
\begin{equation}
{}_sS^{(1)}_{lm}(\theta,\phi) = \sum_j b^\prime_{jl} \; {}_sS^{(0)}_{jm}(\theta,\phi),
\end{equation}
into \eqref{eq:firstorder}. In this way, we can obtain the correction $\lambda_1$ to the separation constant and the coefficients $b^\prime_{jl}$ of the expansion by applying the usual theory of perturbative expansion. There are terms of the form:
\begin{equation}
\int d\Omega\; {}_sY^{*}_{lm}(\theta,\phi) \sin^{n_1}(\theta)\cos^{n_2}(\theta)\; {}_sY_{lm}(\theta,\phi)
\end{equation}
which can be treated as in section \ref{sec:small} by writing the sine and cosine functions as superpositions of spherical harmonics or spin-weigthed spherical harmonics and then using formulae \eqref{eq:pro1} - \eqref{eq:profin}.

We find, for negative spin:
\begin{equation}
\lambda_{1} = \frac{2a M m s}{\sqrt{\lambda_{0}}} - \frac{4 M a \omega m s}{\sqrt{\lambda_{0}}} \frac{l(l+1)(-1+l+l^2+m^2)+s^2(l^2+l-3m^2)}{l(l+1)(2l-1)(3+2l)} - \frac{aMm}{\sqrt{\lambda_{0}}} \left( 1+\frac{s^2}{l^2+l} \right).
\end{equation}
In the case of positive spin, we have that: 
\begin{equation}\label{eq:subs}
\lambda_{0} \mapsto \lambda_{0}+1.
\end{equation}

\subsection{Second order}
At second order, we proceed as in the previous subsection using equation \eqref{eq:secondorder}, thus finding:
\begin{equation}
\begin{split}
\lambda_{2} =&\frac{a^2s^2M^2}{\lambda_{0}}\,\frac{9m^2-4sm^2(2l-1)(2l+3)+2(l(l+1)-s^2)(l(l+1)-3m^2)}{l(l+1)(2l-1)(3+2l)}+\\
&+ \frac{maM\lambda_{1}}{\lambda^{3/2}_{0}}\,\left[ \frac{M}{2} \, \left(1+\frac{s^2}{l(l+1)} \right) - s \right] + \frac{2Mams}{\lambda^{1/2}_{0}} c^{\prime\prime}_{lm} +\\
&+\frac{2Mas\, a\omega}{\lambda_{0}} \left( \frac{Mas}{5} + \frac{m\,\lambda_{1}}{3\,\lambda^{1/2}_{0}} \, \frac{l(l+1)(9m^2+l(l+1)-3) + 9s^2\,(l(l+1)-3m^2)}{l(l+1)(2l-1)(3+2l)} \right),
\end{split}
\end{equation}
where we have defined:
\begin{equation}
c^{\prime\prime}_{lm} = \sum_{k\neq l} \frac{\int d\Omega\left[{}_sY^{*}_{lm}(\theta,\phi)(a-a\omega\,\sin^2(\theta))^2\,{}_sY_{km_1}(\theta,\phi)\right]}{l(l+1) - k(k+1)}
\end{equation}
For positive spin, one should again carry out the substitution \eqref{eq:subs}.

\subsection{Special case}
There is a special case in this semi-analytical approach, the one in which:
\begin{equation}
\left\{\begin{array}{ll}
\lambda_0\equiv0 & \text{for }s=-\dfrac{1}{2},\\
\lambda_0+1\equiv0 & \text{for }s=\dfrac{1}{2}.\\
\end{array} \right.
\end{equation}
One can see that, in this case, any of the corrections given in the previous two subsections will diverge, so this case requires special care.

If we substitute \eqref{eq:expang2} with $\lambda_0(+1)=0$ into \eqref{eq:angular}, we see that, for consistency with our expansion of the separation constant, we have to impose $\lambda_1=0$, otherwise we would have corrections at order $O(\tilde{m}_e^{n/2})$ with odd $n$ which are not included into the expansion \eqref{eq:expang2}. Moreover, since in the term
\begin{equation}
\frac{2M a m s\,\tilde{m}_e}{\tilde{m}_e\,(\sqrt{\lambda_2+O(\tilde{m}_e^3)}-2Mams\,\cos(\theta))}
\end{equation}
the $\tilde{m}_e$ present in the denominator will simplify the one present in the numerator, we find that, at \emph{order} $O(\tilde{m}_e^0)$:
\begin{equation}
\begin{split}
&\frac{d^2S_{(0)}^{lms}(\theta,\phi)}{d\theta^2} + \left( \cot(\theta)+\frac{a\,M\,\sin(\theta)}{\sqrt{\lambda_2} -2 \,aMs\,\cos(\theta)} \right)\,\frac{d\;S_{(0)}^{lms}(\theta,\phi)}{d\theta} +\\
&+\left[ L + s - \left( s\,\cot(\theta)+\frac{m}{\sin(\theta)} \right)^2 -a^2\omega^2\,\sin^2(\theta)+2m\,a\omega + \frac{2aMms}{\sqrt{\lambda_2}-2aMms\,\cos(\theta)}\, (1-a\omega\,\sin^2(\theta)) \right]\, S_{(0)}^{lms}(\theta,\phi)=0,
\end{split}
\end{equation}
where:
\begin{equation}
\left\{\begin{array}{ll}
L\equiv 0 & \text{for }s=-\dfrac{1}{2},\\
L\equiv -1 & \text{for }s=\dfrac{1}{2}.\\
\end{array} \right.
\end{equation}
This equation has to be consistent with the zeroth order one, eqn. \eqref{eq:angular2}. This means that we have to impose:
\begin{equation}\label{eq:condlam}
\frac{2aMms}{\sqrt{\lambda_2}-2aMms\,\cos(\theta)}\,\left( \frac{1}{2}\,\frac{d\;S_{(0)}^{lms}(\theta,\phi)}{d\theta} + (1 - a\omega\,\sin^2(\theta))\,S^{lms}_{(0)}(\theta,\phi) \right) = -L.
\end{equation}
This is a condition on $\lambda_2$. We need to multiply the above expression on the left by the conjugate of $S_{(0)}^{lms}(\theta,\phi)$, integrate over the sphere and solve \emph{numerically} the resulting expression for $\lambda_2$. In order to carry out the integration, it is convenient to use the change of variable:
\begin{equation}
\cos\left(\frac{\theta}{2}\right) \mapsto x \qquad  \theta \in [0,\pi] \mapsto x\in [0,1]
\end{equation}
and use the properties of the spin-weighted spherical harmonics reported in the previous section.

Finally, as an aside, we notice that in this special case, in order not to have terms at order $O(\tilde{m}_e^{n/2})$ with odd $n$, we have to impose that \emph{for any odd} $i$, $\lambda_i\equiv 0$, so that only even order corrections are needed in $\lambda$ in this case.

We now turn to the problem of solving the radial Teukolsky equation \eqref{eq:radial}.

\section{Solution of the radial Teukolsky equation}
\label{sec:radial}

\subsection{General properties of the radial Teukolsky equation}
In this section, we look for the solution of the radial Teukolsky equation \eqref{eq:radial}. Because of the presence of the fermion mass $\tilde{m}_e$, the solution of the equation is much more complicated.

We see that, at variance with respect to the \emph{usual} mass-less case, in \eqref{eq:radial} there are four singularities:
\begin{itemize}
\item the two singularities at finite $r$, already present in the usual treatment, which are located at the horizons of the black hole and are given by the solution of $\Delta=0$, i.e.:
\begin{equation}
r_\pm= M \pm \sqrt{M^2-a^2} = M \pm M \sqrt{1-q^2}
\end{equation}
where $a= q\, M$ is the reduced angular momentum of the black hole (as in \cite{lrr});
\item the singularity at $r=\infty$, also already present in the usual treatment, and finally,
\item a singularity at:
\begin{align}
&r_0= S_n = \frac{\sqrt{\lambda}}{2i M \tilde{m}_e s} & \text{for negative spin,}\\
&r_0= S_n = \frac{\sqrt{\lambda+1}}{2i M \tilde{m}_e s} &\text{for positive spin.}
\end{align}
\end{itemize}
The presence of the fourth singularity complicates the solution, which should be sought for in the form of Heun functions, whose differential equation is given by (see for example \cite{nist}):
\begin{equation}\label{eq:heunode}
H^{\prime\prime}(x) + \left( \frac{\gamma}{x} + \frac{\delta}{x-1} + \frac{\eta}{x-A} \right) H^\prime(x) + \frac{-Q +x \alpha \beta}{x(x-1)(x-A)} H(x)=0
\end{equation}
with:
\begin{equation}\label{eq:parcond}
\eta=\alpha+\beta+1-\gamma-\delta,
\end{equation}
while $Q$ is a number called the \emph{accessory parameter}. Another possibility could be Stieltjes polynomials (see again \cite{nist}), which are generalizations of Heun functions and are given by the solutions of the differential equation:
\begin{equation}
G^{\prime\prime}(x) + \left( \frac{\gamma}{x} + \frac{\delta}{x-1} + \frac{\eta}{x-A} \right) G^\prime(x) + \frac{V(x)}{x(x-1)(x-A)} G(x)=0,
\end{equation}
where $V(x)$ is called \emph{van Vleck potential} and, in this case, is a polynomial of degree 1; for Stiltjes polynomials there is no relation between the parameters $\gamma$, $\delta$ and $\eta$. However, as we shall discuss below, we have encountered some difficulties in finding a recurrence relation for Heun functions, so we employ the method described in \cite{nist,expansion,expansion2} which permits to expand an Heun function in terms of a set of Gauss' Hypergeometric functions ${}_2F_1(a,b,c;x)$. Thus the result will be similar to the usual case, albeit more complicated from the mathematical point of view. Moreover, as in the \emph{usual} case, the solution will not be convergent in the whole domain $r_{+}<r<\infty$, but, as explained also in references \cite{nist,expansion,expansion2}, there will be two separated solutions: one inside an ellipse $\mathcal{E}$ passing through the \emph{new} singularity at $r_{0}$ and with the foci at the position of the other two finite singularities,  and a solution converging outside $\mathcal{E}$. However, matching the two solutions is easier in this case than in the \emph{usual} one \cite{nist,expansion,expansion2}.

\subsection{The general solution at the horizon}

In order to find a solution for the radial Teukolsky equation, we first make the definitions:
\begin{equation}
\kappa=\sqrt{1-q^2} \quad \epsilon = 2 \omega M \quad \tau = \frac{\epsilon - q m}{\kappa} \quad A=\frac{M-\kappa M-S_n}{2\kappa M} \quad \omega x= \frac{r_+-r}{\epsilon \kappa}
\end{equation}
As usual, we use Frobenius method, by assuming that the solution can be expanded in series around each finite singularity as follows:
\begin{equation}\label{eq:ind}
{}_sR(x) = \exp(i\epsilon \kappa x)\,x^g\, \sum_n \alpha_n\,x^n,
\end{equation}
where $\alpha_n$ are some coefficients. We can find the index $g$ of the solution by substituting the above expression into \eqref{eq:radial}, imposing $n=0$ and looking at the coefficient of the lowest power of $x$ of the resulting expression; we find:
\begin{equation}
(i\,\epsilon+2g+2s+i \tau)(\epsilon+2i\,g+\tau)=0,
\end{equation}
which has two solutions:
\begin{equation}
g_1=-s-\frac{i}{2}(\epsilon+\tau) \qquad g_2=\frac{i}{2}(\epsilon+\tau);
\end{equation}
only the first gives a converging solution, so this is the index $g$ appearing in \eqref{eq:ind}. We can proceed in an analogous way for the other two singularities at $x=1$ and $x=A$. We find that the indices for these two singularities are, respectively:
\begin{equation}
g = \frac{i}{2}(\epsilon-\tau) \qquad g = 0.
\end{equation}
As a result, we see that the solution to the radial Teukolsy equation has the form:
\begin{equation}
{}_sR(x) = \exp(i \kappa \epsilon x)(-x)^{-s-\frac{i}{2}(\epsilon+\tau)}(1-x)^{\frac{i}{2}(\epsilon-\tau)} \, S_l(x).
\end{equation}
We notice that this has the same form of the solution for the mass-less case, see \cite{lrr,libro}.

The function $S_l(x)$ is the solution of the differential equation:
\begin{equation}\label{eq:o1}
\frac{d^2S_l}{dx^2} + \left( \frac{\gamma}{x} + \frac{\delta}{1-x}  + \frac{\eta}{A-x} \right) \frac{dS_l}{dx} + \frac{V(x)  S_l(x)}{x(1-x)(A-x)} = 0,
\end{equation}
where we have imposed:
\begin{equation}\label{eq:par}
\gamma=1-s -i \tau + i \epsilon \quad \delta=1+s-i\tau -i \epsilon \quad \eta= 1,
\end{equation}
while the potential $V(x)$ is given by:
\begin{equation}\label{eq:potential}
\begin{split}
V(x)&= (A-x)\,\Big[ \lambda+s(s+1)+\tau(i+\tau)-M^4\,\tilde{m}_e(1+\kappa(1-2x))^2 +\\
&+ \epsilon \, \Big( i\,\kappa(1-2s+2(-1+s+i \tau)) \Big) + \epsilon^2\,\Big( 1-2\kappa x \Big) \Big] + \frac{i\,\epsilon}{2} \, \Big( 1+2\kappa(x-1)x+s(2+4(-1+\kappa(x-1))x) \Big).
\end{split}
\end{equation}
Neglecting terms with powers of $x$ larger than 1, we can impose, using \eqref{eq:parcond}:
\begin{align}
\alpha&=1-i\,\tau - \Big[1+\lambda+s(s+1)-i\,\tau -3\, M^4\,\tilde{m}_e\,(1-\kappa)^2 + 4\,(1+\kappa)(1+\kappa-2A\,\kappa)\,M^4\,\tilde{m}_e^2 +\\
&+ 2i\,\epsilon \Big( s+ \kappa + A\,\kappa(1-s -i\tau) \Big) - \epsilon^2\,(1+2A\,\kappa) \Big]^{1/2}\\
\beta&=1-i\,\tau - \Big[1+\lambda+s(s+1)-i\,\tau -3\, M^4\,\tilde{m}_e\,(1-\kappa)^2 + 4\,(1+\kappa)(1+\kappa-2A\,\kappa)\,M^4\,\tilde{m}_e^2 +\\
&+ 2i\,\epsilon \Big( s+ \kappa + A\,\kappa(1-s -i\tau) \Big) - \epsilon^2\,(1+2A\,\kappa) \Big]^{1/2}\\
Q&=-A\,\Big( \lambda + s(s+1)+\tau(\tau+i) - M^4\,\tilde{m}_e - \epsilon\, i\, \kappa\,(1-2s) + \epsilon^2\Big) -\frac{i\,\epsilon}{2}(1+2s).
\end{align}

If we also define the two functions:
\begin{align}
V_2(x) &= x^2\, \Big[ -4\,\kappa\,M^4\,\tilde{m}_e^2\,\big( 1+\kappa + \kappa\, A \big) + \epsilon \, \kappa\,(3i+2\,\tau) -2\kappa \Big],\\
V_3(x) &=4\,\kappa\,M^4\,\tilde{m}_e^2\,x^3,
\end{align}
we see that we can rewrite \eqref{eq:o1} as follows:
\begin{equation}\label{eq:TeuHeu}
\frac{d^2S_l}{dx^2} + \left( \frac{\gamma}{x} + \frac{\delta}{1-x}  + \frac{\eta}{A-x} \right) \frac{dS_l}{dx} + \frac{-Q +x\, \alpha\,\beta}{x(1-x)(A-x)}\,S_l(x) + \frac{V_2(x)+V_3(x)}{x(1-x)(A-x)}\,S_l(x) = 0,
\end{equation}
from which it is apparent the similarity of this equation and the Heun differential equation \eqref{eq:heunode}. This similarity suggests that we may use an approach similar to the usual one, but using Heun functions instead of Hypergeometric functions in order to expand the solution, i.e., to write:
\begin{equation}\label{eq:heunsum}
S_l(x) =\sum_{n=-\infty}^{\infty}f_n H_{\nu+n}(x).
\end{equation}
with $H_{n+\nu}(x)$ an Heun function. However, we have not been able to find a recurrence relation for the $H_{n+\nu}(x)$, so we have resorted to the other approach introduced above, i.e. we expand $S_l(x)$ in terms of a set of Hypergeometric functions following references \cite{nist,expansion}. In some sense, we are assuming that we have found the solution in terms of a sum of Heun functions as in \eqref{eq:heunsum}, expand the result into Gauss' Hypergeometric functions and finally resumming the various terms.

We now discuss this approach.

\subsection{Heun functions and their expansion in terms of Gauss' Hypergeometric functions}
\label{sec:heun}
As recalled above, Heun functions, $H(x)$, are solutions of the Heun differential equation, which has the general form \cite{nist}:
\begin{equation}
\begin{split}
&H^{\prime\prime}(x) + \left( \frac{\gamma}{x} + \frac{\delta}{x-1} + \frac{\eta}{x-A} \right) H^\prime(x) + \frac{-Q +x \alpha \beta}{x(x-1)(x-A)} H(x)=0,\\
&\alpha+\beta+1-\gamma-\delta=\eta.
\end{split}
\end{equation}

It is generally possible to expand a Heun function into a sum of Hypergeometric functions ${}_2F_1(a,b,c;x)$, where, using Riemann's notation \cite{nist,expansion}:
\begin{equation}
F_n(x)={}_2F_{1}(\sigma+n,\mu-n,\gamma;x)= F\left\{ \begin{array}{cccc}
0 & 1 & \infty &\\
0 & 0 & \sigma+n & x\\
1-\gamma & 1-\delta & \mu-n &
\end{array} \right\}.
\end{equation}

As explained in \cite{nist,expansion,expansion2} this expansion can be carried out in many different ways, which can be grouped into two classes: \emph{type I} and \emph{type II}. For type I expansions, one fixes either:
\begin{equation}\label{eq:lambdamu1}
\sigma=\alpha \quad \mu=\beta-\eta,
\end{equation}
or
\begin{equation}
\sigma=\beta \quad \mu=\alpha-\eta.
\end{equation}
Type I expansions converge outside the ellipse $\mathcal{E}$ defined above.

For type II expansions we have four possibilities:
\begin{alignat}{2}
&\sigma=\gamma+\delta+1 & \mu=0,\\\label{eq:lambdamuusata}
&\sigma=\gamma & \mu=\delta-1,\\
&\sigma=\delta & \mu=\gamma-1,\\\label{eq:lambdamu2}
&\sigma=1 & \mu=\gamma+\delta-2.
\end{alignat}
Type II expansions converge inside $\mathcal{E}$.

The Hypergeometric functions $F_n(x)$ satisfy the diferrential equation:
\begin{equation}
\frac{d^2F_n}{dx^2} + \left( \frac{\gamma}{x} + \frac{\delta}{x-1} \right) \,\frac{dF_n}{dx} + (\lambda +n)(\mu -n) \,F_n(x) = 0.
\end{equation}
If we multiply the above by $(x-A)$ and subtract the result to \eqref{eq:heunode} multiplied by $x(x-1)(A-x)$, we find:
\begin{equation}\label{eq:resto2}
\eta\,x(x-1)\, \frac{dF_n}{dx} + (-Q + \alpha\, \beta \, x- (\sigma+n)(\mu-n))\, F_n(x) = 0.
\end{equation}
We now need to find a recurrece relation for the $F_n$. We first notice that from the properties of the Hypergeometric functions we have \cite{nist}:
\begin{equation}\label{eq:hyperder}
x(x-1)\,F'_n(x)= ND^{(1)}_{n-1}F_{n-1}(x) + ND^{(2)}_n F_n(x) + ND^{(3)}_{n+1} F_{n+1}(x),
\end{equation}
where:
\begin{subequations} \label{eq:def_ND}
\begin{equation}
ND^{(1)}_{n-1}=-\frac{(\mu-n)(\sigma+n)(\sigma+n-\gamma)}{\sigma-\mu-1+2n)(\sigma-\mu+2n)},
\end{equation}
\begin{equation}
ND^{(2)}_{n}=\frac{(1-2\gamma+\sigma+\mu)(\mu-n)(\sigma+n)}{(\sigma-1-\mu+2n)(\sigma+1-\mu+2n)},
\end{equation}
\begin{equation}
ND^{(3)}_{n+1}=\frac{(\mu-n)(\sigma+n)(\gamma-\mu+n)}{(\sigma-\mu+2n)(1+\sigma-\mu+2n)}.
\end{equation}
\end{subequations}
From the series expansion of the Hypergeometric function, we can also prove that:
\begin{equation} \label{eq:xrecu}
x\,F_n(x)= N^{(1)}_{n-1}F_{n-1}(x) + N^{(2)}_n F_n(x) + N^{(3)}_{n+1} F_{n+1}(x),
\end{equation}
where:
\begin{subequations} \label{eq:def_N}
\begin{equation}
N^{(1)}_{n-1}=\frac{(\mu-n)(\sigma-\gamma+n)}{(\sigma+2n-\mu-1)(\sigma+2n-\mu)},
\end{equation}
\begin{equation}
N^{(2)}_{n}=\frac{(\mu-1)\,\gamma+2n\,(n-\mu)+(\sigma+n)\,(\gamma-2\mu+2n)}{(\sigma+2n-1-\mu)(1+2n+\sigma-\mu)},
\end{equation}
\begin{equation}
N^{(3)}_{n+1}=\frac{(\mu-\gamma-n)(\sigma+n)}{(\sigma-\mu+2n)(1+\sigma-\mu+2n)}.
\end{equation}
\end{subequations}
By substituting repeatedly \eqref{eq:hyperder} - \eqref{eq:def_N} into \eqref{eq:resto2}, we can find the recurrence relation we were looking for.

We now come back to the Teukolsky equation.

\subsection{Expansion of the Teukolsky equation in terms of Hypergeometric functions}

In order to find a solution at the horizon converging inside the ellipse $\mathcal{E}$, we use an expansion of type II, in particular, we use the definition for $\sigma$ and $\mu$:
\begin{equation*}
\sigma = \gamma+\delta-1 \qquad \mu =0;
\end{equation*}
see \cite{expansion} for the procedure for determining the correct values of the expansion parameter $\sigma, \mu.$

By following the procedure outlined above, we are left with the equation:
\begin{equation}\label{eq:vera}
x(x-1)\, \frac{dF_n}{dx} + (-Q + \alpha\, \beta \, x + V_2(x) + V_3(x)  - (\gamma+\delta+n-1)(-n))\, F_n(x) = 0.
\end{equation}
Now, following \cite{lrr}, we sum and subtract the quantity $[\nu(\nu+1)-\lambda-s(s+1)]\,F_n(x)$ in the above expression and essentially follow the usual treatment described in \cite{lrr}: we treat the term $[V_2(x)+V_3(x)]\,F_n(x)$ as a perturbation with small parameters $\epsilon$ and $\tilde{m}_e$, look for a formal solution of the form:
\begin{equation}\label{eq:expa}
S_l(x) = \sum_{n=-\infty}^{\infty} f_n \, F_n(x)
\end{equation}
and finally impose the condition that the series must converge, thus finding an (implicit) equation for the renormalized angular momentum $\nu$.

Following the procedure outlined in the previous subsection, we find that the resulting recurrence relation has the form:
\begin{equation}\label{eq:recu}
\alpha^\nu_{0,n}a_{n}+\alpha^\nu_{1,n}a_{n+1}+\alpha^\nu_{2,n}a_{n+2}+\alpha^\nu_{3,n}a_{n+3}+\alpha^\nu_{4,n}a_{n+4}+\alpha^\nu_{5,n}a_{n+5}+\alpha^\nu_{6,n}a_{n+6}=0
\end{equation}
The various coefficients $\alpha^\nu_{i,n}$, $i=\{0, \dots ,6\}$ are listed in the appendix \ref{app:uno}. We see that the recurrence relation we have found is not of the three-terms type, but that it has \emph{seven terms}, instead. Not everything is lost, however, since there is a large literature on higher terms recurrence relations and their link to n-continued fractions which we can use, see for example references \cite{recu1,recu2,recu3,recu4,recu5}. 

In the following subsection, we review the mathematics we need for the case in hand.

\subsection{Higher order recurrence relations and n-continued fractions}

An higher order recurrence relation is a recurrence relation with $k > 3$ terms, i.e.:
\begin{equation}\label{eq:recu2}
\alpha^\nu_{0,n}a_{n}+\alpha^\nu_{1,n}a_{n+1}+\alpha^\nu_{2,n}a_{n+2}+\alpha^\nu_{3,n}a_{n+3}+\alpha^\nu_{4,n}a_{n+4}+\dots+\alpha^\nu_{k-1,n}a_{n+k-1}=0
\end{equation}
It can be shown that to \eqref{eq:recu2} is associated an n-continued fraction, i.e. a generalization of continued fractions with more than two parameters. We have also the result that this n-continued fraction has k independent solutions $f^{(1)},\cdots,f^{(k-1)}$ and $h$ such that:
\begin{equation}
\lim_{n\rightarrow\infty} \frac{f^{(i)}_n}{h_n}=0 \qquad i=\{1,\dots,k-1\}
\end{equation}
if the determinant of the matrix
\begin{equation}
\left(\begin{array}{ccc}
f_1^{(1)} & \cdots & f_{k-1}^{(1)}\\
\cdots & \cdots & \cdots\\
f_{1}^{(k-1)} & \cdots & f_{k-1}^{(k-1)}
\end{array}\right)
\end{equation}
is different from zero. The $k-1$ solutions $f^{(i)}$ are called \emph{minimal} solutions and are said to be dominated by the solution $h$. We have also the result that a n-continued fraction and its associated recurrence relation are convergent if and only if the all of the $k-1$ convergents of the n-continued fraction have a finite limit, where the i-th convergent is given by:
\begin{equation}
C_n^{(i)}=\frac{A^{(i)}_n}{B_n} \quad i=1,\cdots,k-1
\end{equation}
where $A^{(i)}$ with $(i=1,\cdots,k-1)$ and $B$ are solution of the recurrence relation \eqref{eq:recu2} with the initial conditions:
\begin{align}
&A_j^{(i)}=\delta_{ij} \quad i=1,\cdots,k \quad j=1,\cdots,k+1\\
&B_j=0 \quad j=1,\cdots,n \quad  B_{k+1}=1.
\end{align}
There are different ways to calculate the solutions $A^{(i)}$ and $B$, one of them is to use a matricial approach, which works as follows (see for example \cite{recu1}). We introduce the two $k \times k$ matrices:
\begin{equation}
\mathcal{A}_k=\left( \begin{array}{ccc}
A_j^{(k-1)} & \cdots & A_{j-k+1}^{(n)}\\
\vdots & \ddots & \vdots \\
A_j^{(1)} & \cdots & A_{j-k+1}^{(1)}\\
B_j & \cdots & B_{j+1-k}\\
\end{array} \right)
\end{equation}
\begin{equation}
\mathcal{B}_k =\left( \begin{array}{cccc}
b_j& 1 & \cdots &0\\
a_j^{(k-1)} & 0 & 1 & \cdots\\
\cdots & \cdots & \cdots & \cdots\\
a_j^{(1)} & \cdots & \cdots & 0
\end{array} \right)
\end{equation}
then the $j+1$ term of the solutions is given by $\mathcal{A}_{j+1}=\mathcal{A}_{j} \mathcal{B}_{j+1}$. The starting matrix $\mathcal{A}_0$ is given by:
\begin{equation}
\mathcal{A}_0 = \left(\begin{array}{cccc}
a_0^{(k-1)} & 1 & \cdots & 0\\
a_0^{(k-2)} & 0 & \cdots &0\\
\cdots & \cdots & \cdots & \cdots\\
1 & 0 & \cdots & 0
\end{array}\right),
\end{equation}
where $a_0^{(i)}$ are the initial conditions. The convergents of the n-continued fraction can be calculated as follows: we first introduce the operator $s_j$ which takes the vector $\zeta=\{\zeta^{(1)},\cdots,\zeta^{(k)}\}$ into the vector
\begin{equation}
s_j=\left( \frac{a_j^{(1)}}{b_j+\zeta^{(k)}}, \frac{a_j^{(2)}+\zeta^{(1)}}{b_j+\zeta^{(k)}}, \cdots, \frac{a^{(k-1)}_j+\zeta^{(k-1)}}{b_j+\zeta^{(k)}} \right)
\end{equation}
then the convergents are given by:
\begin{equation}
C_n= s_n\cdots s_1(0)_k
\end{equation}
where $(0)_k$ is the zero vector with k components.

If we call $d_{(n)}$ the limit of the n-th convergent and we assume it is finite, a theorem (see \cite{recu1,recu2,recu3}) says that a minimal solution of the recurrence relation and of its n-continued fraction \eqref{eq:recu} is given by:
\begin{equation}\label{eq:minimal}
f^{(i)}_m= \sum_{j=1}^{i} A_m^{(j)} - d_{(i)} B_m \qquad i=\{1,\dots,k-1\},
\end{equation}
while the dominant solution is given by:
\begin{equation}
h_m= \sum_{j=1}^{k-1}\left( \alpha_j A_m^{(j)} \right) + \beta B_m
\end{equation}
where $\alpha_j$ and $\beta$ are (possibly complex) numbers. The inverse of the theorem is also valid (see \cite{recu1,recu2,recu3}): if one is given the minimal solutions, then the recurrence relation and the associated n-continued fraction converge.

We can now go back to the Teukolsky equation.

\subsection{Minimal solutions}
By dividing it by $\alpha^\nu_{0,n}$, we can rewrite the recurrence relation \eqref{eq:recu} in the form:
\begin{equation}
a_n = B_n a_{n-1} + A_n^{(5)} a_{n-2} + A_n^{(4)} a_{n-3} + A_n^{(3)} a_{n-4}+ A_n^{(2)} a_{n-5}+ A_n^{(1)} a_{n-6},
\end{equation}
where:
\begin{equation}
B_n = \frac{\alpha^\nu_{1,n}}{\alpha^\nu_{0,n}}, \quad A_n^{(4)}=\frac{\alpha^\nu_{2,n}}{\alpha^\nu_{0,n}}, \quad A_n^{(3)}=\frac{\alpha^\nu_{3,n}}{\alpha^\nu_{0,n}}, \quad A_n^{(2)}=\frac{\alpha^\nu_{4,n}}{\alpha^\nu_{0,n}}, \quad A_n^{(1)}=\frac{\alpha^\nu_{5,n}}{\alpha^\nu_{0,n}}
\end{equation}
Using the procedure described in the above subsection, we see that all the convergents have a finite limit given by:
\begin{align}
d_{(5)}=&\lim_{n\rightarrow\infty}\frac{A^{(5)}_n}{B_n} = -8-4\dfrac{T_2}{T_{12}}-4\dfrac{T_2}{T_3},\\
d_{(4)}=&\lim_{n\rightarrow\infty}\frac{A^{(4)}_n}{B_n} = 13+24 \dfrac{T_2}{T_{12}} +8 \dfrac{T_2}{T_3} +16\, \dfrac{T_{02}T_2}{T_{12}T_3},\\
d_{(3)}=&\lim_{n\rightarrow\infty}\frac{A^{(3)}_n}{B_n} = -6-4\dfrac{T_2}{T_3},\\
d_{(2)}=&\lim_{n\rightarrow\infty}\frac{A^{(2)}_n}{B_n} = 0,\\
d_{(1)}=&\lim_{n\rightarrow\infty}\frac{A^{(1)}_n}{B_n} = 0,
\end{align}
where the coefficients are reported in the appendix \ref{app:uno}. By the theorem recalled above, the n-continued fraction and the associated recurrence relation converge and a linear combination of the minimal solutions \eqref{eq:minimal} gives the coefficients $f_n$ in the expansion \eqref{eq:expa}.

\subsection{Outer solution and matching}

As reported in references \cite{expansion,expansion2} and also in \cite{nist}, there is a relation between type I and type II expansions for the same differential equation. Starting from an expansion of type II with the form like the one we used above converging inside $\mathcal{E}$:
\begin{equation}\label{eq:expin}
H(x)= \sum_n A_n \;{}_2F_1(\gamma + \delta + n -1, - n ,\gamma,x)
\end{equation}
we have that a solution \emph{of the same differential equation} converging outside $\mathcal{E}$ can be found by solving the integral:
\begin{equation}
F(x) = \Lambda\,\left( 1-\frac{x}{A} \right)^{\eta-1} \, \int_0^1\, d\zeta\, \zeta^{\gamma-1}\,(1-\zeta)^{\delta-1}\;{}_2F_1(\gamma + n,\delta - n - 1,\gamma,x)\; {}_2F_1\left(\alpha-\eta,\beta+1-\eta,\gamma;\frac{x\,\zeta}{A}\right).
\end{equation}
The result of the integration is reported in cite{expansion}, and it is given by:
\begin{equation}\label{eq:expout}
\begin{split}
H(x) &= \Lambda \sum \Big[ (-1)^n  \dfrac{\Gamma(\alpha+n)\Gamma(\beta+n)\Gamma(\delta+n)}{\Gamma(\gamma+n)\Gamma(\gamma+\delta+2n)} \; \left( \dfrac{x}{A} \right)^n \, {}_2F_1\left(\alpha+n,\beta+n,\gamma+\delta+2n; \dfrac{x}{A}\right) \Big]
\end{split}
\end{equation}
which is of type I. The functions $\Gamma$ are the usual gamma functions \cite{nist}.

In order to fix the parameter $\Lambda$ appearing in equation \eqref{eq:expout}, one needs to match the two inner and outer equation at the point $A$.

What is left to be done now, is to find the expression for the renormalized angular momentum $\nu$: this is done in the following subsection.

\subsection{Determination of $\nu$}

Since there are 5 solutions to the given recurrence relation, there are also 5 different opertors $R^{(i)}_n, \, L^{(i)}_n$ defined in an analogous way as in the usual treatment, i.e.:
\begin{equation}
R^{(i)}_n = \dfrac{f^{(i)}_n}{f^{(i)}_{n-1}} \qquad L^{(i)}_n = \dfrac{f^{(i)}_n}{f^{(i)}_{n+1}} \qquad i=\{1,\dots,5\},
\end{equation}
Similarly to the usual case, the condition 
\begin{equation}\label{eq:cond}
R^{(i)}_nL^{(i)}_{n-1}=1 \qquad i=\{1,\dots,5\}
\end{equation} 
must be enforced for each $i$; as in \cite{lrr}, this is an implicit condition for the parameter $\nu$, but this means that, in priciple, in our case there will be 5 different corrections $\nu^{(i)}$; as a matter of fact, however, only $\nu^{(1)}$ contributes to order $O(\epsilon^2,\tilde{m}_e^2)$. 

Equation \eqref{eq:cond} must be treated differently according to which regime we are considering:
\begin{enumerate}
\item regime $\tilde{m}_e\ll 1$ and $\epsilon\ll 1$: this can be treated by expanding the renormalized angular momentum in series as follows:
\begin{equation}
\nu^{(i)}=l + \tilde{m}_e \nu_{0,1} + \tilde{m}_e^2 \; \nu_{0,2} + \epsilon^2 \; \nu_{2,0} + \tilde{m}_e \epsilon \; \nu_{1,1} +O(\tilde{m}_e^3,\epsilon^3).
\end{equation}
We need to substitute this expression into \eqref{eq:cond}, using the result of section \ref{sec:small} for the expression of the separation constant $\lambda$. We collect the various orders of $\epsilon$ and $\tilde{m}_e$ and set to zero their coefficients, thus finding (four) equations that fix the various $\nu_{i,j}$ defined above; we find that:
\begin{equation}
\nu_{0,1} \equiv 0, \qquad \nu_{0,2} \equiv 0,  \qquad \nu_{1,1}  \equiv 0,
\end{equation}
while the expression for $\nu_{2,0}$ is in general different from zero. This expression is however very long, and we do not report it here, but we make available upon request the Mathematica notebook used for the calculation containing also the results.


\item regime $\tilde{m}_e\ll 1$ and $\epsilon\gtrsim 1$: this case can be treated in a semi-analytical way. We can expand $\nu$ as follows:
\begin{equation}
\nu^{(i)} = \nu_0 + \tilde{m}_e\,\nu_1+ \tilde{m}_e^2 \, \nu_2 +O(\tilde{m}_e^3),
\end{equation}
where $\nu_0$ is the value of the renormalized angular momentum obtained numerically from the \emph{usual} theory for $\tilde{m}_e=0$. We then substitute the above expression into \eqref{eq:cond} and use the results of section \ref{sec:semi-an} for $\lambda$, thus finding that:
\begin{equation}
\nu_1 \equiv 0, \qquad \nu_2 \equiv 0.
\end{equation}
\end{enumerate}

\section{Conclusion}
\label{sec:conclusion}

We have presented a way to solve the Teukolsky equations for a fermion with mass $m_e$ and a rotating black hole of mass $M$. The main difference of these equations with respect to the mass-less case, or with respect to the cases of spin 0, 1 and 2 fields, is the presence of terms depending on the fermion mass $m_e = M\, \tilde{m}_e$ that do not permit to treat the general problem as an eigenvalue problem, as usually done. In our treatment, we have identified four regimes in which the equations can be solved with different approaches:
\begin{enumerate}
\item In the first regime, we have $\tilde{m}_e,a\omega \ll 1$; we have found that it is possible to treat this problem perturbatively by expanding the equations, the separation constant $\lambda$ and the renormalized angular momentum $\nu$ in a double series with small parameters $\tilde{m}_e$ and $\epsilon$. With this approach, the problem can be reduced to an eigenvalue problem, so we can employ a method similar to the usual one, but (unfortunately) much more complicated, since it involves several technical properties of Heun and Gauss' Hypergeometric functions and because of the appearence of a 6-terms recurrence relation. However, we have been able to work through all the needed algebra, finally arriving to analytical corrections to $\lambda$ and $\nu$ up to order $O(\tilde{m}^2_e,\epsilon^2)$, included. We have also given a review of the mathematics involved in the analysis, but for a more detailed and in-depth treatment of this branch of mathematics we refer to the cited references \cite{nist,expansion,expansion2,ip1,ip2,recu1,recu2,recu3,recu4,recu5}.
\item In the second regime, we have, on the contrary, $\tilde{m}_e,a\omega \gtrsim 1$. In this case it is not possible to find a closed form solution, neither for the angular or the radial equation, so we have to resort to a full numerical approach. This regime shall be treated in detail in a forthcoming paper.
\item In the third regime, we have $a\omega\gtrsim 1$ and $\tilde{m}_e\ll 1$. This regime is the one which is probably the most interesting from the astrophysical point of view, since for any known black hole of mass $M$ we have, indeed, $m_e \ll M$. In this case we have used a semi-analytical approach: we have, in fact, calculated numerically with usual methods the values of $\nu$ and $\lambda$ for $\tilde{m}_e=0$ and derived the analytical corrections for a finite fermion mass both for the separation constant $\lambda$ and the renormalized angular momentum $\nu$ up to order $O(\tilde{m}_e^2)$ included. However, there is a draw-back with this method; in fact, in case $\lambda_0=0$ (for negative spin) or $\lambda_0+1=0$ (for positive spin), one has to numerically solve complicated integrals in order to find the corrections to the separating constant $\lambda$.
\item We have also identified a fourth regime, opposite to the last one, in which $a\omega\ll 1$ and $\tilde{m}_e\gtrsim 1$. This regime can also be very interesting, since it might give an hint on the behavior of very small back holes (which might be originated in the very final stage of the Hawking evaporation process, for example) interacting with fermions. However, the treatment of this regime is not reported here, but the full analysis will be presented in a forthcoming, dedicated paper.
\end{enumerate}

\appendix
\section{Recurrence relations coefficients}
\label{app:uno}
In this appendix we report the full expression of the coefficients of the recurrence relations \eqref{eq:recu}.
\begin{equation*}
\alpha^\nu_{0,n+\nu} = T_3\, N^{(1)}_{n-2}\, N^{(1)}_{n-1}\, N^{(1)}_{n},
\end{equation*}
\begin{equation*}
\begin{split}
\alpha^\nu_{1,n+\nu} &= T_2+T_3 \, \Big( N^{(2)}_{n-2} + N^{(2)}_{n-1} + N^{(2)}_n \Big)\, N^{(1)}_{n-1}\,N^{(1)}_n,
\end{split}
\end{equation*}
\begin{equation*}
\begin{split}
\alpha^\nu_{2,n+\nu} &= N^{(1)}_n\, \Big[ T_1 + n\, \Big( T_{11} + n\, T_{12} \Big) + T_3\,(N^{(2)}_{n-1})^2 + T_2\, N^{(2)}_n + T_3\, (N^{(2)}_n)^2 +\\
&+ N^{(2)}_{n-1} \,\Big( T_2 +T_3\, N^{(2)}_n \Big) + T_3 \Big( N^{(3)}_{n-2} N^{(1)}_{n-1} + N^{(3)}_{n-1} N^{(1)}_{n} + N^{(3)}_{n} N^{(1)}_{n+1} \Big) \Big] + ND^{(1)}_n,
\end{split}
\end{equation*}
\begin{equation*}
\begin{split}
\alpha^\nu_{3,n+\nu} &= T_0 + n\,\Big( T_{01} + n  T_{02} \Big) + T_2\, (N^{(2)}_n)^2 + T_3\,(N^{(2)}_n)^3 + ND^{(1)}_n + \Big( T_2 + T_3\, N^{(2)}_{n-1} \Big) \, N^{(3)}_{n-1}\,N^{(1)}_n +\\
&+ \Big( T_2 + T_3 \,N^{(2)}_{n+1} \Big) \, N^{(3)}_n\, N^{(1)}_{n+1} + N^{(2)}_n, \Big[ T_1 + n \, \Big( T_1 + n \, T_{11} + n^2\, T_{12} + 2 T_3\,N^{(3)}_{n-1}\,N^{(1)}_n + 2 \, T_3\, N^{(3)}_n\, N^{(1)}_{n+1} \Big)  \Big],
\end{split}
\end{equation*}
\begin{equation*}
\begin{split}
\alpha^\nu_{4,n+\nu} &= ND^{(3)}_n + N^{(3)}_n\, \Big[ T_1 + n\, T_{11} + n^2\, T_{12} + T_3\, (N^{(2)}_n)^2 + T_2 \, N^{(2)}_{n+1} + T_3\, (N^{(2)}_{n+1})^2 + N^{(2)}\, \Big( T_2 + T_3 \, N^{(2)}_{n+1} \Big) +\\
&+ T_3\, \Big( N^{(3)}_{n-1}\,N^{(1)}_n + N^{(3)}_n\,N^{(1)}_{n+1} + N^{(3)}_{n+1}\,N^{(1)}_{n+2} \Big) \Big],
\end{split}
\end{equation*}
\begin{equation*}
\begin{split}
\alpha^\nu_{5,n+\nu} &= \Big[ T_2 +T_3\, \Big( N^{(2)}_n + N^{(2)}_{n+1} + N^{(2)}_{n+2} \Big)  \Big] \, N^{(3)}_n\, N^{(3)}_{n+1},
\end{split}
\end{equation*}
\begin{equation*}
\begin{split}
\alpha^\nu_{6,n+\nu} &= T_3\, N^{(3)}_n\, N^{(3)}_{n+1}\, N^{(3)}_{n+2}.
\end{split}
\end{equation*}

The expressions for $N^{(i)}_n$ are given in eqn. \eqref{eq:def_N}, while the definitions of the coefficients appearing in the above expressions are given below:
\begin{equation*}
T_0 = \dfrac{1}{4}\, \Big[ 1+\, \gamma(4+\gamma) - (\delta-2) + 4\, A\, \Big( \lambda - \sigma \mu-(1+\kappa)^2M^4 \tilde{m}_e^2 + (1+\gamma)\delta + i\, \epsilon\, (\delta-1-\gamma + \kappa, (\delta+\gamma-1)) \Big) \Big],
\end{equation*}
\begin{equation*}
T_{01} = A\,( \sigma-\mu ),
\end{equation*}
\begin{equation*}
T_{02} = A,
\end{equation*}
\begin{equation*}
T_1 = 4\, A\, \epsilon^2\,\kappa-\lambda + \sigma\, \mu + (\kappa+1)(8A\,\kappa-\kappa-1) M^4\, \tilde{m}_e^2 + i \, \epsilon\, \Big[ 2\kappa(-1+A\,\gamma)-1-2\gamma \Big] - \delta (\gamma+1),
\end{equation*}
\begin{equation*}
T_{11} = -\sigma+\mu
\end{equation*}
\begin{equation*}
T_{12} = -1,
\end{equation*}
\begin{equation*}
T_2 = \kappa\, \Big[ -4\, \epsilon^2-4(1+\kappa+A\,\kappa)\,M^4\, \tilde{m}_e^2 - i\, \epsilon\,(\gamma-\delta-1) \Big],
\end{equation*}
\begin{equation*}
T_3 = 4\kappa^2\,M^4\, \tilde{m}_e^2.
\end{equation*}

We also remind the definition of $A$:
\begin{equation*}
A = \left\{\begin{array}{lr}
\dfrac{1}{2} + \dfrac{m\,s\,M^2\,\tilde{m}_e+i\,\sqrt{\lambda}}{2m\,s\,\kappa\,M^2\,\tilde{m}_e} & s<0,\\
\dfrac{1}{2} + \dfrac{m\,s\,M^2\,\tilde{m}_e+i\,\sqrt{\lambda+1}}{2m\,s\,\kappa\,M^2\,\tilde{m}_e} & s>0.\\
\end{array}\right.
\end{equation*}

\end{document}